\documentclass[proof]{WileyASNA-v1}
\usepackage{tikz}
\usetikzlibrary{mindmap,trees}
\usepackage{url}
\usepackage{graphicx}
\usepackage{color}
\usepackage{bm}

\newcommand{\udt}[3]{#1^{#2}_{\phantom{#2}#3}}

\articletype{Article Type}%

\received{October 2020}
\revised{}
\accepted{}

\begin{document}

\title{Precision Cosmology in Modified and Extended Theories of Gravity: an insightful test}

\author[1]{Celia Escamilla-Rivera}

\authormark{Celia Escamilla-Rivera}

\address[1]{\orgdiv{Instituto de Ciencias Nucleares, Universidad Nacional Aut\'onoma de M\'exico, Circuito Exterior C.U., A.P. 70-543, M\'exico D.F. 04510, M\'exico.}}  

\corres{*Circuito Exterior C.U., A.P. 70-543, M\'exico D.F. 04510, M\'exico. \email{celia.escamilla@nucleares.unam.mx}}

\abstract{In this work we present a brief discussion about modified and extended cosmological models using current observational tests. We show that according to these astrophysical samples 
based in late universe measurements, theories like $f(R)$ and $f(T,B)$ can provide useful interpretation to a dynamical dark energy. At this stage, precision cosmostatistics
has also become a well-motivated endeavour by itself to test gravitational physics at cosmic scales and these analyses 
can be employed to test the viability and future constrains over specific cosmological models of these theories of gravity, making them a good approach to propose an alternative path from the standard $\Lambda$CDM scenario.}


\maketitle

\section{Introduction}

Now days, models that try to explain the late-time behaviour of the $\Lambda$CDM model has been called into question. The major problem is presented as a
form of the so-called $H_0$ tension, which characterises the disparity between late-time model-independent measurements of the expansion of the universe and their corresponding model-dependent predictions from the early times \cite{Aghanim:2018eyx,Ade:2015xua}. A statistical relevance due to strong lensing by the H0LiCOW ($H_0$ lenses in Cosmograil's wellspring) collaboration \cite{Wong:2019kwg} and measurements from Cepheids via SH0ES (Supernovae $H_0$ for equation of state) \cite{2006hst..prop10802R} have reinforced such problematic. While Tip of the Red Giant Branch (TRGB Carnegie-Chicago Hubble Program) measurements have recovered a lower $H_0$ tension value \cite{Freedman:2019jwv}.

To solve (or at least relax) this issue, rather than modifying the matter content of the models, gravity needs to be modified in a way that we can obtain the current dynamics observed. In this line of thought, there has been a variety of proposal in which General Relativity (GR) can be modified or extended \cite{Clifton:2011jh}. One interesting approach has been the $f(R)$ theories \cite{Sotiriou:2008rp,Faraoni:2008mf} (and references therein), where we can consider an arbitrary function of the Ricci scalar in the standard Einstein-Hilbert action. As a consequence of introducing this arbitrary function, there is a freedom to explain the late cosmic acceleration and structure formation without adding any kind of dark energy or dark matter. 

Recently, Teleparallel Gravity (TG) has starting to take advantage of its mathematical description to solve the $H_0$ tension problem and describe the late-time dynamics observed and well-tested with current data. In these kind of theories the curvature description is replaced with a torsion term by the replacement of the Levi-Civita connection with its Weitzenb\"{o}ck connection \cite{Weitzenbock1923}. At this level, GR and TG can be made equal to a boundary term in the Teleparallel equivalent of General Relativity (TEGR). Both can produce identical dynamical field equations \cite{Aldrovandi:2013wha, Blixt:2018znp,Aldrovandi:2004fy}. There are several ways to modify the TEGR proposal, one through the so-called torsion scalar $T$. With this technique, we can generalise the Lagrangian to arbitrary functions of the torsion scalar to produce $f(T)$ gravity \cite{Ferraro:2006jd,Linder:2010py,Chen:2010va}, which follows the same idea as $f(R)$ gravity. However, unlike $f(R)$ theory, $f(T)$ gravity produces generally second-order field equations. This result is interesting since it means that Lovelock's theorem is weakened in TG \cite{Lovelock:1971yv} which has had interesting consequences for constructing scalar-tensor theories of gravity \cite{Bahamonde:2019shr}. Furthermore, given the importance of the boundary term in relating GR and TEGR, $f(T,B)$ gravity has also been well-studied \cite{Escamilla-Rivera:2019ulu,Bahamonde:2015zma,Farrugia:2018gyz} as a possible extension to TEGR. 

Moreover, applying consistency tests with current astrophysical data allows to identify an optimal gravity theory and deal at the same time with systematic effects in the data or any problems with the underlying cosmological model. Some data samples are sensitive to the geometry and dynamics of the universe and some other samples are sensitive to the growth of large-scale structure. In such case, these two sets of observations must be consistent with one another in order to solve the cosmological tension inside an optimal theory of gravity. At late-times, any deviations between theories can be measured through an effective equation of state (EoS) (mimicking a dark energy component) close to the $\Lambda$CDM EoS value ($w=-1$). 

In this work we provide an insightful test via examples in $f(R)$ and $f(T,B)$ gravities that can relax the $H_0$ tension using current late-time surveys. We present a description of both theories to use them (with their corresponding constraints and priors) to study the nature of the effective EoS in both cases. 
Due to the straightforward calculations, sometimes we will be forced to restrict ourselves to displaying only the results and interpret them qualitatively for 
cosmological priors in both of the samples at tension: Planck 2018 and Cefeids measurements. 


\section{Modified/Extended theories of gravity versus dark energy}

GR is based on well-defined principles, most of which, are contained in the structure of the Einstein tensor and their field equations. In other words, from the Lovelock theorem that can be summarised as follows:
\textit{The only second-order, local gravitational field equations derivable from an action containing solely the 4D metric tensor are the Einstein field equations with a cosmological constant.}

\begin{figure}[htb]
\centering
\includegraphics[width=90mm,height=50mm]{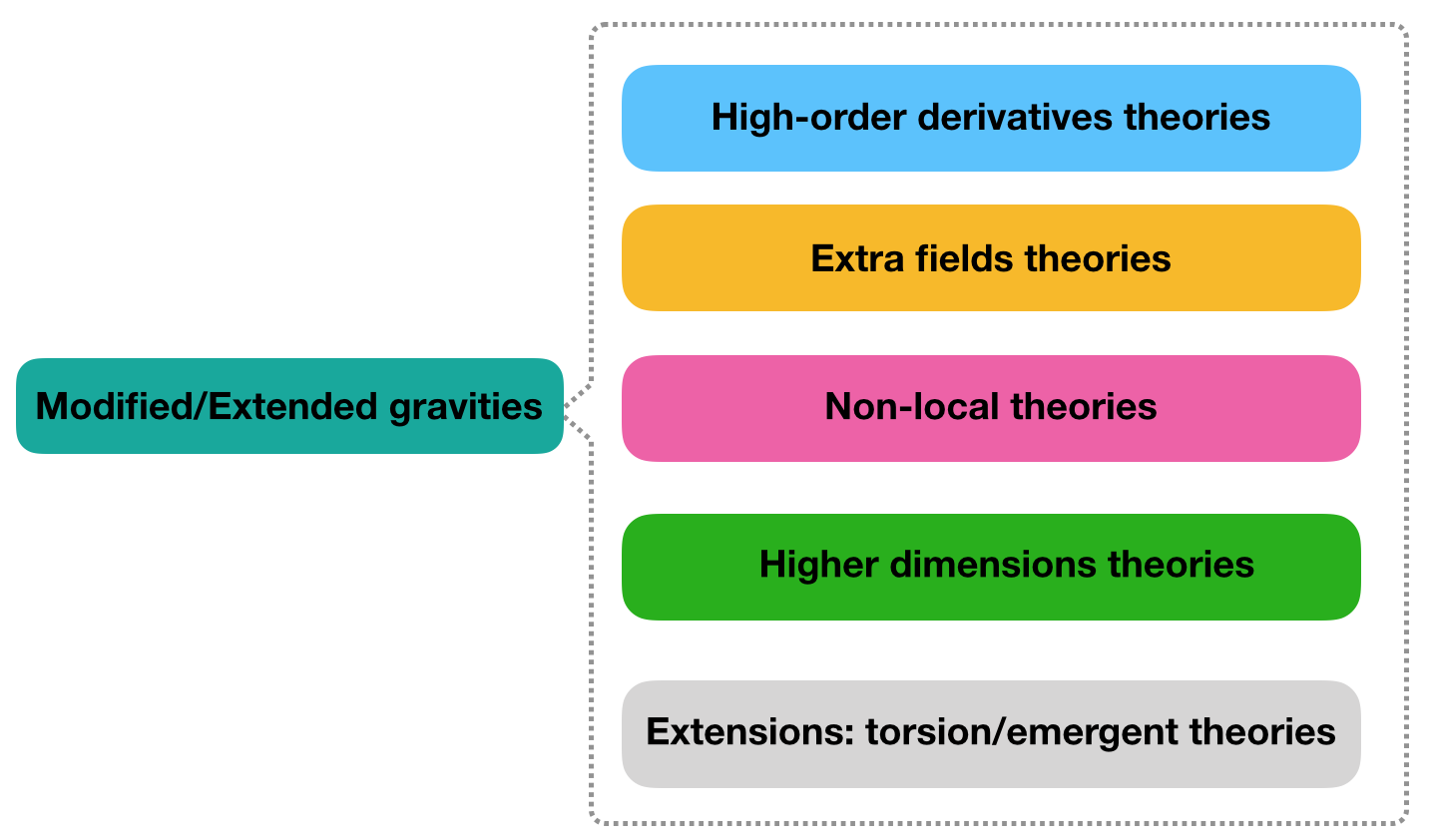} 
\caption{Several categories of modified gravity (MG) theories according to the Lovelock theorem and their requirement of violate it.} 
\label{fig:gravities_categories}
\end{figure}

\begin{figure}[htb]
\centering
\includegraphics[width=90mm,height=50mm]{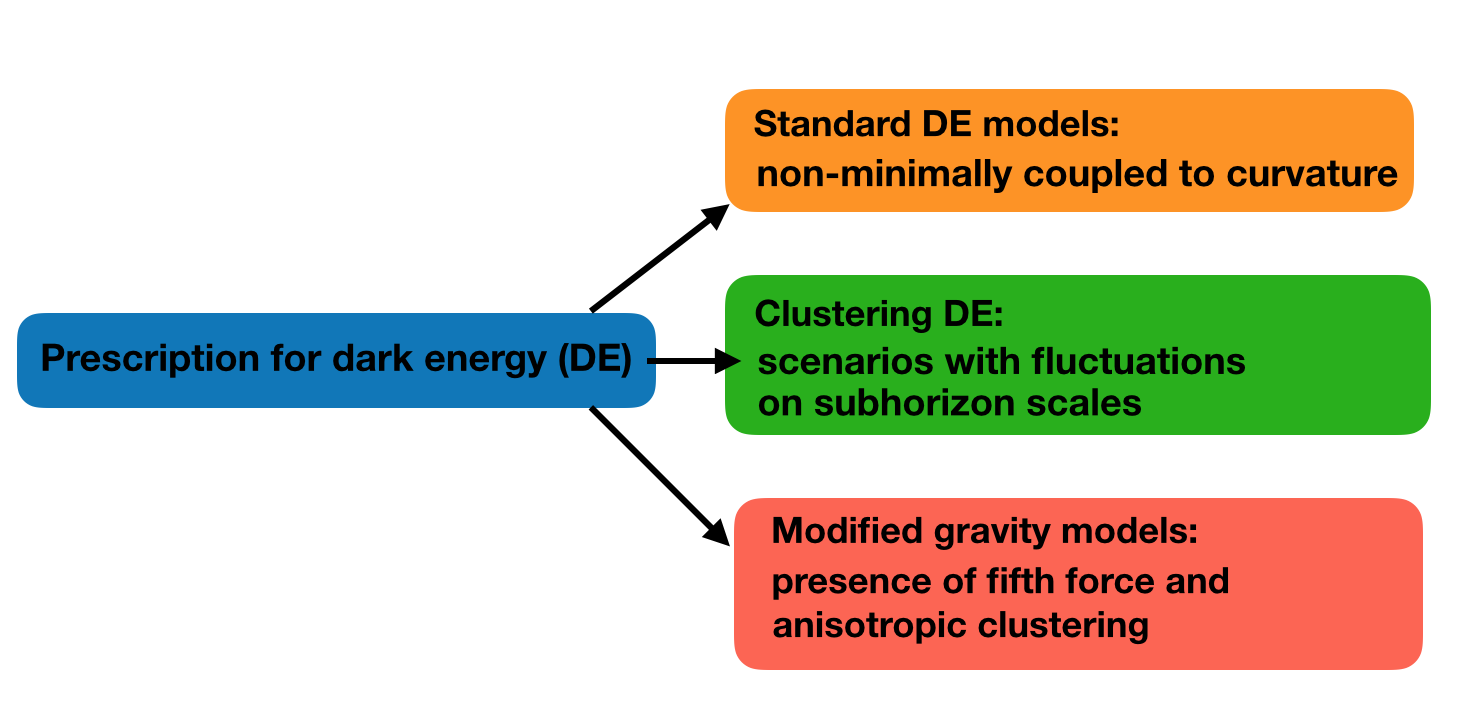} 
\caption{A convenient (phenomenological) prescription for dark energy.} 
\label{fig:DE prescription}
\end{figure}

Theories that deviate from GR can be delineated into categories according to what principle or requirement they violate \cite{Escamilla-Rivera:2020sba}. A modification to GR can thus happen by allowing one of more options described in Fig.~\ref{fig:gravities_categories}. In this line of thought, an optimal prescription can be put down directly into three categories that can describe the effectiveness of the late-time dynamics, in such case mimicking a dark energy component (see Figure \ref{fig:DE prescription}). 

\section{Modified and extended gravities}

\subsection{Illustrative example 1: $f(R)$ theories}

These theories of gravity take a general function of the Ricci scalar in the Einstein-Hilbert action as
\begin{equation}
\label{f(R)}
S[g_{\mu\nu},{\mbox{\boldmath{$\psi$}}}] =
\!\! \int \!\! \frac{f(R)}{2\kappa} \sqrt{-g} \: d^4 x 
+ S_{\rm matt}[g_{\mu\nu}, {\mbox{\boldmath{$\psi$}}}] \; ,
\end{equation}
where $G=c=1$ and $\kappa \equiv 8\pi$, $S_{\rm matt}[g_{\mu\nu}, {\mbox{\boldmath{$\psi$}}}]$ is the standard 
action for matter. And $f(R)$ is an arbitrary function of the Ricci scalar $R$. 

As it is standard, we can consider a flat homogeneous, isotropic universe described by the Friedman-Lema\^itre-Robertson-Walker (FLRW) metric
and with the variation we can derive the equations

\begin{eqnarray}
\label{traceRt}
& \ddot R = &-3H \dot R -  \frac{1}{3 f_{RR}}\left[ 3f_{RRR} \dot R^2 + 2f- f_R R + \kappa T \right], \,\,\,\,\,\, \\
\label{Hgen}
& H^2 = & -\frac{1}{f_{RR}}\left[f_{RR}H\dot{R}-\frac{1}{6}(Rf_{R}-f) \right]-\frac{\kappa T^{t}_{t}}{3f_{R}}, \\
\label{Hdotgen}
& \dot{H}= & -H^2 -\frac{1}{f_{R}} \left[ f_{RR}H\dot{R} + \frac{f}{6}+\frac{\kappa T^{t}_{t}}{3} \right]  \,\,\,,
\end{eqnarray}
where $H = \dot a/a$. The energy momentum tensor $T$ is that for a fluid composed by baryons, dark matter and radiation. 
According to \cite{Jaime:2018ftn}, 
by integrating these equations using a fourth order Runge-Kutta numerical integration, we can derive 
the evolution for $R$, $H$ and its derivative to obtain a generic EoS given by:
\begin{equation}
\label{eq:fit}
\omega_{\text{JJE}} = -1 + \frac{w_0}{1+w_1z^{w_2}}cos(w_3+z).
\end{equation}
where $\omega_i$ are free parameters and $z$ is the redshift given by $z=a_0/a-1$. This is the so-called JJE parameterisation (\cite{Jaime:2018ftn,Escamilla-Rivera:2020giy}). This EoS recovers $\omega=-1$ at large $z$ and allows dynamical oscillation. In this work we will consider models that can provide an accelerated evolution, with a $\omega_X \approx -1$, e.g the Starobinksy model $ f(R)= R+\lambda R_{S}\left[ \left( 1+\frac{R^2}{R^2_{S}}\right)^{-q}-1\right]$ with $\lambda=0.9$,   $R_S=1$\footnote{These values were taken from observational Solar System test \cite{Jaime:2018ftn}.}. We shall call this case as $f(R)$ power law.

\subsection{Illustrative example 2: $f(T,B)$ theories}
To study a cosmology that emerges from $f(T,B)$ gravity, we consider a flat homogeneous and isotropic metric. We choose to take again the FLRW metric $
ds^2=-dt^2+a(t)^2(dx^2+dy^2+dz^2)\,,$
where the lapse function is set to unity. This can be done since $\tilde{f}(T,B)$ gravity retains diffeomorphism invariance. If we consider the choice of tetrad as
$
\udt{e}{a}{\mu}=\mbox{diag}(1,a(t),a(t),a(t))\,,$
the spin connection components are set to be zero, i.e. $\omega^{a}{}_{b\mu}=0$ \cite{Bahamonde:2016grb}. In these theories can exist an infinite number of possible choices for the $\udt{e}{a}{\mu}$, but only a small set have vanishing associated spin connection components. The torsion scalar and the  boundary term are given by
\begin{equation}\label{torsionscalar_frw}
T = 6H^2\,, \quad\quad  B = 6\left(3H^2+\dot{H}\right)\,,
\end{equation}
with them, we can write the Ricci scalar as
\begin{equation}
R = -T+B = 6\left(\dot{H} + 2H^2\right)\,,
\end{equation}
This shows how $f(R)$ gravity results as a subset of $f(T,B)$ gravity with
\begin{equation}
f(T,B) := \tilde{f}(-T+B) = \tilde{f}(R)\,.
\end{equation}
Taking a standard arbitrary Lagrangian mapping 
$\tilde{f}(T,B) \rightarrow -T + f(T,B)\,,$
and for an universe filled with a perfect fluid we 
we can obtain \begin{eqnarray}
\kappa^2 \rho_{\mbox{eff}} &=& 3H^2\left(3f_B + 2f_T\right) - 3H\dot{f}_B + 3\dot{H}f_B - \frac{1}{2}f\,, \label{eq:friedmann_mod}\\
\kappa^2 p_{\mbox{eff}} &=& \frac{1}{2}f-\left(3H^2+\dot{H}\right)\left(3f_B + 2f_T\right)-2H\dot{f}_T+\ddot{f}_B\,, \quad\quad
\end{eqnarray}
with
$2\dot{H}=-\kappa^2\left(\rho_m + p_m + \rho_{\mbox{eff}} + p_{\mbox{eff}}\right).$
An EoS can be computed for this effective fluid as
\begin{eqnarray}
w_{\text{TG}} 
&=& -1+\frac{\ddot{f}_B-3H\dot{f}_B-2\dot{H}f_T-2H\dot{f}_T}{3H^2\left(3f_B+2f_T\right)-3H\dot{f}_B+3\dot{H}f_B-\frac{1}{2}f}\,. \label{EoS_func} \quad\quad
\end{eqnarray}
First, as in (\ref{eq:fit}), notice here that we can again recover the $\Lambda$CDM scenario when the $T$ and $B$ terms are negligible. Second, this EoS require a specific form
of $f(T,B)$. In analogy to $f(R)$, in this work we consider a power law expression of the form $f(T,B) = b_0 B^k + t_0 T^m$,
to obtain the effects of a late-time cosmic accelerated expansion without the influence of a dark energy component.

\section{Late-time cosmological surveys}

\subsection{Cosmic Chronometers}

This sample consist in passively evolving old galaxies whose redshifts are known. As an advantage, the expansion history of the universe can be inferred directly from their differential ages. According to this, we consider the current data discussed in Table \ref{tab:cc}.

\begin{table}[h]
\centering
\begin{tabular}{lllllllll}
\hline
$z$    & $H(z)$ & $\sigma_{H(z)}$ & $\quad$ & $z$   & $H(z)$ & $\sigma_{H(z)}$  \\
\hline
0.07   & 69.0 & 19.6        & & 0.4783 & 80.9   & 9.0   \\
0.09   & 69.0 & 12.0        & & 0.48   & 97.0   & 62.0  \\
0.12   & 68.6 & 26.2       & & 0.593  & 104.0  & 13.0 \\
0.17   & 83.0 & 8.0       & & 0.68   & 92.0   & 8.0      \\
0.179  & 75.0 & 4.0   & & 0.781  & 105.0  & 12.0  \\
0.199  & 75.0 & 5.0      & & 0.875  & 125.0  & 17.0  \\
0.2    & 72.9 & 29.6         & & 0.88   & 90.0   & 40.0\\
0.27   & 77.0 & 14.0      & & 0.9    & 117.0  & 23.0  \\
0.28   & 88.8 & 36.6      & & 1.037  & 154.0  & 20.0  \\
0.352  & 83.0 & 14.0     & & 1.3    & 168.0  & 17.0 \\
0.3802 & 83.0 & 13.5     & & 1.363  & 160.0  & 33.6  \\
0.4    & 95.0 & 17.0     & & 1.43   & 177.0  & 18.0  \\
0.4004 & 77.0 & 10.2    & & 1.53   & 140.0  & 14.0  \\
0.4247 & 87.1 & 11.2     & & 1.75   & 202.0  & 40.0  \\
0.4497 & 92.8 & 12.9   & & 1.965  & 186.5  & 50.4  \\
0.47   & 89.0 & 49.6&  & &  \\
\hline
\end{tabular}
\caption{Cosmic Chronometers data sample from  \cite{Marra:2017pst}.}
\label{tab:cc}
\end{table}


\subsection{Type Ia Supernovae}

The current sample in this category is the compressed supernova Ia Pantheon compilation (40 bins)\footnote{\url{https://github.com/dscolnic/Pantheon}}.
The standard description provide numerical values of the distance modulus $\mu$, whose can be directly employed to derive the luminosity distance $d_{L}$ (in Mpc) according to:
\begin{equation}
\mu\left(z\right)=5\log\left[\dfrac{d_{L}\left(z\right)}{1 \text{ Mpc}}\right]+25\,.
\label{mu}
\end{equation} 
Usually, we add to this quantity the nuisance parameter $M$, which is related (and degenerate) to the prior on $H_{0}$. Therefore, in this work we will consider additional priors to calibrate this sample. Assuming spatial flatness, $d_{L}$ is related to the comoving distance $\mathcal{D}$ as
\begin{equation}
\mathcal{D}\left(z\right)=\dfrac{H_{0}}{c}\left(1+z\right)^{-1}10^{\frac{\mu\left(z\right)}{5}-5}\,,
\end{equation}
which can be normalised by the Hubble function $E\left(z\right)\equiv H\left(z\right)/H_{0}$ to obtain
$\mathcal{D}\left(z\right)=\int_{0}^{z}\dfrac{d\tilde{z}}{E\left(\tilde{z}\right)}$.

\subsection{Baryon Accoustic Oscillations}

For our analyses, also we will consider model-independent angular  BAO 
determinations from the angular correlation function. We use a total of 
14 uncorrelated data points (see Table~\ref{tab-bao}).
\begin{table}[t]
\centering
\begin{tabular}{lllll}
\hline
Catalog & $z$    & $\theta(z)$ & $\sigma_{\theta(z)}$ \\
\hline
SDSS-DR7          & 0.235        & 9.06        & 0.23        \\
SDSS-DR7          & 0.365        & 6.33        & 0.22      \\
SDSS-DR10         & 0.450        & 4.77        & 0.17     \\
SDSS-DR10         & 0.470        & 5.02        & 0.25      \\
SDSS-DR10         & 0.490        & 4.99        & 0.21       \\
SDSS-DR10         & 0.510        & 4.81        & 0.17     \\
SDSS-DR10         & 0.530        & 4.29        & 0.30    \\
SDSS-DR10         & 0.550        & 4.25        & 0.25      \\
SDSS-DR11         & 0.570        & 4.59        & 0.36      \\
SDSS-DR11         & 0.590        & 4.39        & 0.33       \\
SDSS-DR11         & 0.610        & 3.85        & 0.31        \\
SDSS-DR11         & 0.630        & 3.90        & 0.43        \\
SDSS-DR11         & 0.650        & 3.55        & 0.16       \\
SDSS-DR12Q$\quad$ & 2.225$\quad$ & 1.77$\quad$ & 0.31$\quad$  \\
\hline
\end{tabular}
\caption{BAO data sample (only angular) from \cite{deCarvalho:2017xye}.}
\label{tab-bao}
\end{table}

The theoretical BAO angular scale $\theta\left(z\right)$ can be written using the angular diameter distance $d_{A}\left(z\right)$, which, for a flat universe is  
$\mathcal{D}\left(z\right)$,
\begin{equation}
\mathcal{D}\left(z\right)=\dfrac{H_{0}}{c}\dfrac{r_{s}}{\theta\left(z\right)}\left(\dfrac{180}{\pi}\right)\,.
\label{Dbao}
\end{equation}
where $r_{s}$ is the sound horizon of the primordial photon-baryon fluid .

\subsection{$H_0$ and $M$ priors} 

The first value under consideration come from model-independent measurements of the local determination of 
$H_0$ obtained from low-redshift SN Ia data calibrated with local Cepheids \cite{Riess:2018byc}.
Our second value it's from the current Planck 2018
\cite{Aghanim:2018eyx} obtained with 
TT,TE,EE+lowE+lensing+BAO.
Since $H_{0}$ is fixed by both values, we have nuisance parameters $M$, which can be calculated by calibrating the Pantheon sample described with each of them. The results of the statistical analysis at 1-$\sigma$ C.L is shown in Table~\ref{tab-M}. 

\begin{table}[t]
\centering
\begin{tabular}{|c|c|c|c|} 
 \hline 
 Prior & $H_{0}$ & $M$ & $r_{s}$  \\ \hline 
Riess et al.$\quad$ & $73.52$ &  $-19.25^{+0.01}_{-0.01}$ & $146.6\pm4.1$  \\ 
Planck 2018 & $67.66$ & $-19.42^{+0.01}_{-0.01}$ & $147.21\pm0.23$\\ 
\hline
 \end{tabular} \\ 
\caption{Result of the statistical analysis  calibration for the Pantheon sample.}
\label{tab-M}
\end{table}

Additionally, when using the angular BAO determinations, we have $r_{s}$ as an extra parameter. Again, using both $H_0$ values we can compute the values reported in the fourth column in Table~\ref{tab-M}. 

\section{Performing Precision Cosmology at large scales}

\begin{figure}[htb]
\centering
\includegraphics[width=47mm,height=37mm]{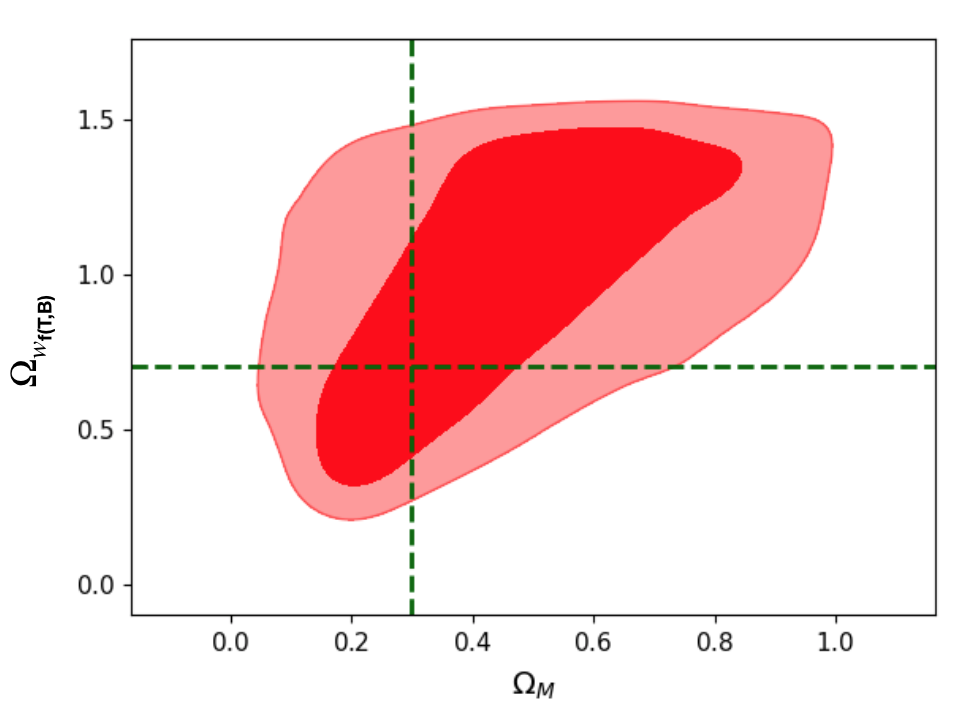} 
\includegraphics[width=47mm,height=37mm]{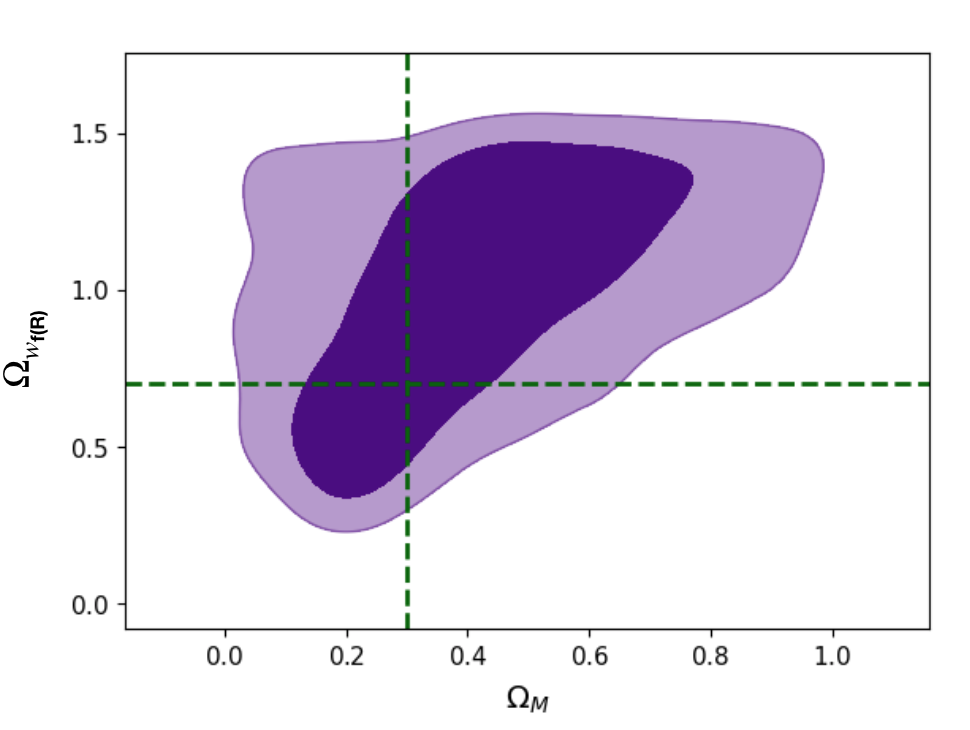} 
\caption{$f(R)$ cosmological power law models. \textit{Top:} with Planck 2018 prior. \textit{Bottom:} with Riess et al prior.} 
\label{fig:FRmodel}
\end{figure}

\begin{figure}[htb]
\centering
\includegraphics[width=47mm,height=37mm]{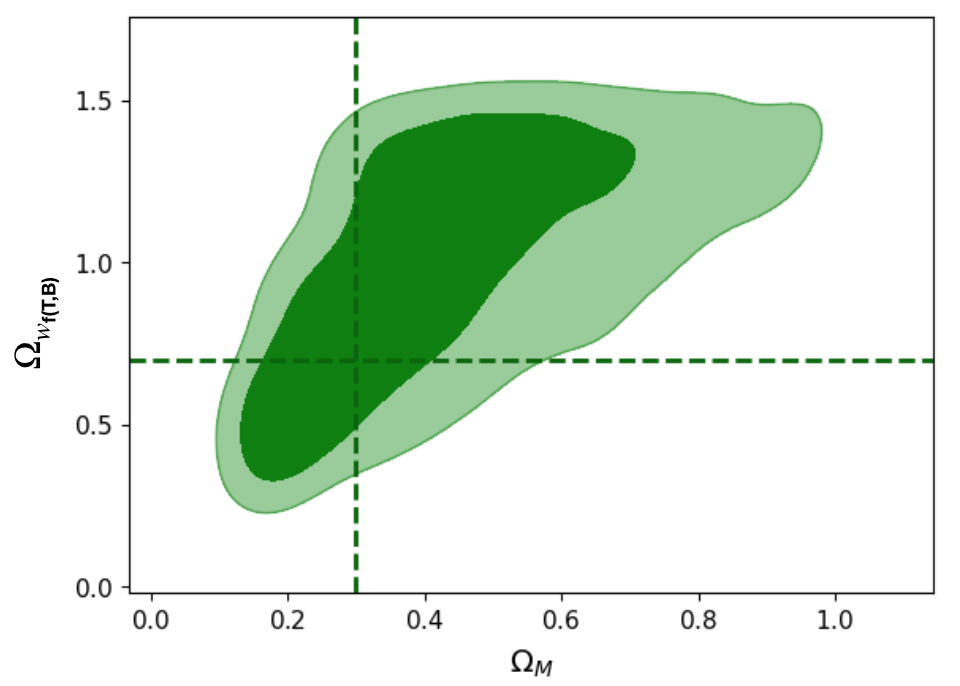} 
\includegraphics[width=47mm,height=37mm]{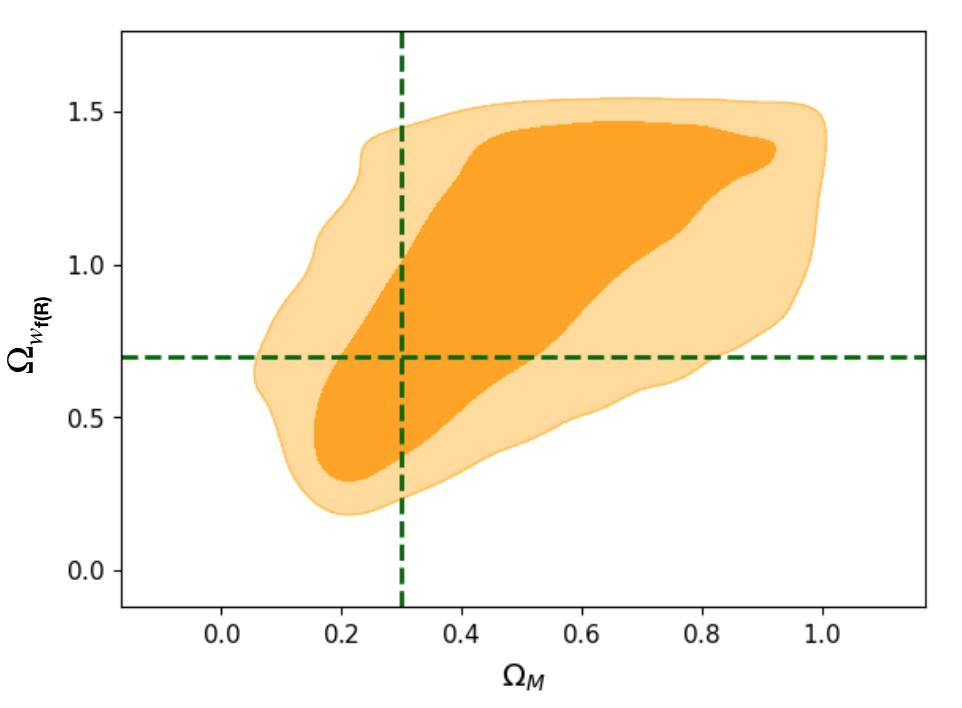} 
\caption{$f(T,B)$ cosmological power law models. \textit{Top}: with Planck 2018 prior. \textit{Bottom:} with Riess et al prior.} 
\label{fig:FTBmodels}
\end{figure}

\begin{figure}[htb]
\centering
\includegraphics[width=74mm,height=53mm]{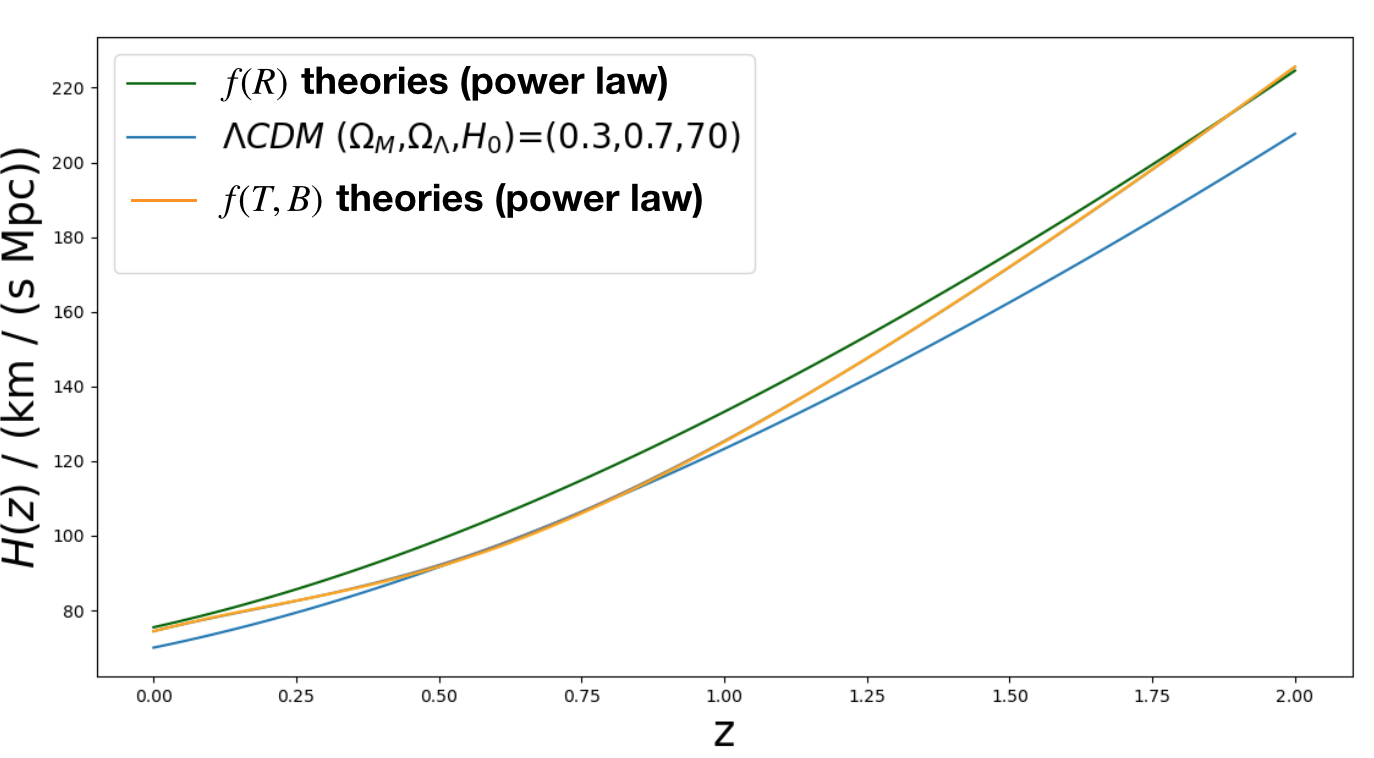} 
\caption{$f(R)$ and $f(T,B)$ cosmological power law models in comparison to the Benchmark $\Lambda$CDM model ($\Omega_m=0.3$, $h=0.7$ and $H=70$).} 
\label{fig:all_models}
\end{figure}

Using a modified version of the MontePython code in order to introduce our EoS's for $f(R)$ (\ref{eq:fit}) and $f(T,B)$ (\ref{EoS_func}) power law models. To compare the parameterisations with a Benchmark $\Lambda$CDM we use the combination of the three different datasets (CC+Pantheon+BAO) with the $H_0$ and $r_s$ priors and calculate the corresponding bestfit for the late-time cosmic acceleration effect. Finally, we obtained the following 
main results:
\begin{itemize}
\item Power Law $f(R)$ Model: Using our total sample we obtain the mean constraints $w_0=0.6961_{-6.0524 e^{-03}}^{+6.0524 e^{-03}}$, $w_1 =0.03244_{-0.0203}^{+0.0203}$, $w_2=0.1188_{-8.1688 e^{-04}}^{+8.1688 e^{-04}}$ and $w_3=2.246_{-1.2451 e^{-04}}^{+1.2451 e^{-04}}$.
In this scenario there is no notable difference between the $H_0$ priors employed (see Figure \ref{fig:FRmodel}). Moreover, a deviation of 1-$\sigma$ is obtained in comparison to a Benchmark model (see Figure \ref{fig:all_models}).
\item Power Law $f(T,B)$ Model: When our total sample is used, we obtain the mean constraints $m=79.19_{-6.1}^{+4}$, $49.81_{-1}^{+0.73}$, $b_0= 1.099e+16_{-1.1e+16}^{+7e+14}$ and $t_0 = 7.974e+15_{-6.3e+15}^{+2.8e+15}$. It seems to be in agreement with the value $H_0$ given by Riess et al (see Figure \ref{fig:FTBmodels}).
This model mimic a $\Lambda$CDM at low $z$, while at higher value starts to reproduce a $f(R)$ power law (see Figure \ref{fig:all_models}). 
\end{itemize}

\section{Conclusions}

In this work we discussed a straightforward cosmological analysis for $f(R)$ and $f(T,B)$ theories described by generic EoS expressions in a flat homogeneous and isotropic space-time. As a first approach, we proposed $f(R)$ and $f(T,B)$ power law cosmological scenarios.
On the theoretical level, $f(T,B)$ acts as a generalisation of $f(R)$ gravity in which the torsion and boundary term contributions are decoupled from each other. For a flat FLRW space-time, this means that some viable $f(R)$ models can allow for more freedom. Moreover, it would be interesting to use the tools employed for an extended theories of gravity to probe interesting additions from the decoupled scalars to approximate different phenomena in late-time cosmology. This study will be reported elsewhere.

\bigskip
\textit{\bf Acknowledgments.-}
CE-R acknowledges networking support by the COST Action CA18108, Royal Astronomical Society as FRAS 10147 and 
PAPIIT Project IA100220.
\bibliography{Template}%

\end{document}